\documentclass[prl,twocolumn,superscriptaddress]{revtex4}
\usepackage{t1enc}
\usepackage{amsmath}
\usepackage{graphicx}

\begin{document}
\newcommand{\kb}{k_\mathrm{B}^{{}}}
\newcommand{\eF}{\varepsilon_\mathrm{F}^{{}}}
\newcommand{\ern}{\varepsilon_\mathrm{R}^{{}}}
\newcommand{\er}{\varepsilon_\mathrm{R}}
\newcommand{\e}{\varepsilon}
\newcommand{\kfa}{k_{\mathrm{F}a}^{{}}}
\newcommand{\kfb}{k_{\mathrm{F}b}^{{}}}
\newcommand{\TFa}{T_{\mathrm{F}a}^{{}}}
\newcommand{\TFb}{T_{\mathrm{F}b}^{{}}}
\newcommand{\vfa}{v_{\mathrm{F}a}^{{}}}
\newcommand{\vfb}{v_{\mathrm{F}b}^{{}}}
\newcommand{\vfat}{v_{\mathrm{F}a}}
\newcommand{\dd}{\mathrm{d}}
\newcommand{\lee}{l_{\textrm{e}\textrm{e}}}
\newcommand{\mub}{\mu_{\textrm{B}}^{}}

\title{Interaction-Induced Resonance in Conductance and Thermopower of Quantum Wires}
\author{Anders Mathias Lunde}
\affiliation{Nano-Science Center, Niels Bohr Institute, University
of Copenhagen, DK-2100 Copenhagen, Denmark} \affiliation{William I.
Fine Theoretical Physics Institute, University of Minnesota,
Minneapolis, MN 55455, USA}

\author{Karsten Flensberg}
\affiliation{Nano-Science Center, Niels Bohr Institute, University
of Copenhagen, DK-2100 Copenhagen, Denmark}

\author{Leonid I. Glazman}
\affiliation{William I. Fine Theoretical Physics Institute, University of
Minnesota, Minneapolis, MN 55455, USA}
\date{\today}
\begin{abstract}
  We study the effect of electron-electron interaction on the
  transport properties of short clean quantum wires adiabatically
  connected to reservoirs. Interactions lead to resonances in a
  multi-channel wire at particular values of the Fermi energy. We
  investigate in detail the resonance in a two-channel wire. The
  (negative) conductance correction peaks at the resonance, and decays
  exponentially as the Fermi energy is tuned away; the resonance width
  being given by the temperature. Likewise, the thermopower shows a
  characteristic structure, which is surprisingly well approximated by
  the so-called Mott formula.  Finally, four-fold splitting of the
  resonance in a magnetic field provides a unique signature of the
  effect.
\end{abstract}

\maketitle

It is well established by now that ballistic motion of electrons
in quantum wires results in the conductance
quantization~\cite{wees1988,Wharam1988}. With the increase of wire
width, the conductance increases by one quantum, $2e^2/h$, each
time a new channel becomes available for the electron propagation.
The experimental observation of the low-temperature conductance
quantization was explained successfully within the model of
non-interacting electrons~\cite{glazman88}.  In this model,
corrections to the quantized conductance values come from the
electron diffraction at the edges of the wire and from thermal
broadening. At the conductance plateaus the latter corrections are
exponentially small at low temperatures.

There is little reason to believe that the electron-electron
interactions are weak in the studied quantum wires. Nevertheless,
apart maybe from the so-called ``$0.7$ anomaly'', they apparently do
not show up in the experimental observations of quantized steps in
the dependence of conductance on the wire width. The lack of the
effect of interaction on the low-temperature conductance at the
first plateau can be easily understood within the Luttinger liquid
picture~\cite{maslov} and from the fact that the density-density
interaction does not re-distribute electrons between the left moving
and right moving species.

Within the model of non-interacting electrons, the thermopower is
related to the derivative of the electron transmission coefficient
with respect to energy in the vicinity of the Fermi level [giving
rise to the so-called Mott's formula, see Eq.~\eqref{eq:mott-law}
below]. Therefore such a model predicts zero thermopower at the
conductance plateaus in good agreement with
experiments~\cite{exp,vanhouten92,appleyard00}. Again, interactions,
if accounted for within the Luttinger liquid approximation, does not
alter this result due to the particle-hole symmetry built into the
approximation.

\begin{figure}
\centerline{\includegraphics[width=0.2\textwidth]{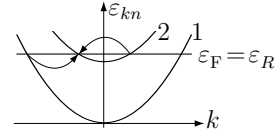}} \caption{An
example of inter-channel scattering event discussed in
  this paper. This scattering event becomes possible at the Fermi
  level only if the ratio of the Fermi momenta of the two channels
  equals $3:1$.\label{fig:fig1}}
\end{figure}

In this paper, we find interaction induced features in the electron
transport of a multi-channel wire. The features have the form of
resonances in the dependence of conductance and thermopower on the
Fermi energy (or, equivalently, on the gate voltage). The origin of
the features is in the possibility of interaction-induced electron
scattering between the channels at some specific relations of the
Fermi wave vectors of electrons in different channels. These
relations are determined by energy and momentum conservation. The
inter-channel scattering events do not necessarily preserve the
number of right- and left-movers. For example, in
Fig.~\ref{fig:fig1} a right and a left mover become two left movers.
Since the number of right (left) movers is not conserved in this
scattering event, it changes the particle current. We evaluate in
detail the corrections to the conductance $G$ and the thermopower
$S$ for a two-channel wire due to such processes.

\begin{figure}
\includegraphics[width=0.35\textwidth]{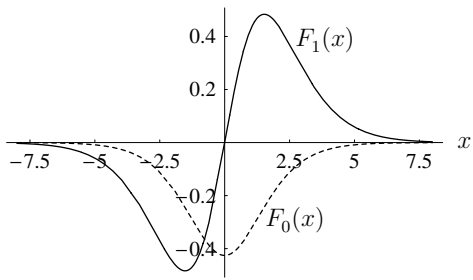}
\caption{The dimensionless scaling functions $F_0(x)$ and $F_1(x)$
entering the conductance correction Eq.~(\ref{eq:conductance}) and
the interaction-induced thermopower Eq.~(\ref{eq:thermopower}),
respectively. \label{fig:fig2}}
\end{figure}

We show that near the resonance point displayed in
Fig.~\ref{fig:fig1}, the interaction-induced correction to the
quanized value of conductance, $4e^2/h$, has the form
\begin{equation}
\delta G=\frac{4e^2}{h}\frac{L}{\lee}\frac{\kb T}{\eF}
F_0\left(\frac{\eF-\ern}{\kb T}\right),
\label{eq:conductance}
\end{equation}
where $\er=(9/8)\varepsilon_0$, with $\varepsilon_0$ being the
difference of the quantized energies of the transverse motion
corresponding to the two channels, 1 and 2, see Fig.~\ref{fig:fig1};
$L$ is the wire length, and $\lee$ is the electron mean free path
corresponding to the described type of scattering. We assume weak
scattering, so the wire is sufficiently short, $L\ll(\eF/\kb
T)\lee$, see Eq.~(\ref{eq:leedef}) below. The dimensionless function
$F_0(x)$ (given in Fig.~\ref{fig:fig2} and Eq.~\eqref{eq:Fn}) is of
the order of unity when $|x|\lesssim 1$ and falls off exponentially
at large $|x|$. The wire width and thus the difference $\eF-\ern$
can be controlled electrostatically~\cite{wees1988,Wharam1988} by a
gate voltage.  The shape of the conductance dip can be easily
understood from the process in Fig.~\ref{fig:fig1}. For an excess of
right-movers created by the bias voltage, it allows for a relaxation
of the right-movers into the left-moving species.

\begin{figure}
\centerline{\includegraphics[width=0.4\textwidth]{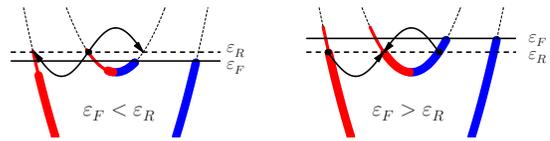}}
\caption{(Color online) Illustration of the dominant scattering
  events on either side of the resonance, $\eF=\ern$, for a finite
  temperature difference across the wire. Here the right movers are
  cold (blue) and left movers are warmer (red); for simplicity we
  consider $\mu_{\textrm{L}}=\mu_{\textrm{R}}$ and $T=0$ for right
  movers. Due to momentum and energy conservation the scattering has
  to take place close to $\ern$ (the dotted line). In the left panel,
  only scattering of electrons from warm to cold is possible and
  consequently an excess of right-movers is created.  In contrast, in
  the right panel this process is blocked by a filled Fermi sea and
  instead a scattering where a right mover is transformed to a left
  mover prevails. This explains the sign change of thermopower seen in
  Fig.~\ref{fig:fig2}.\label{fig:fig3}}
\end{figure}

When the conductance is quantized at the value $4e^2/h$, the
thermopower vanishes in \textit{absence} of interactions. Accounting
for the interactions, we find the leading contribution to the
thermopower coefficient $S$ for two open channels
\begin{equation}
S=\frac{\kb}{e}\frac{L}{\lee}\frac{\kb T}{\eF}F_1
\left(\frac{\eF-\ern}{\kb T}\right). \label{eq:thermopower}
\end{equation}
The thermopower $S$ is proportional to $T$ at the maximum. Here $S$
is conventionally defined by the relation $\Delta V=S\Delta T$, with
$\Delta T$ and $\Delta V$ being, respectively, the temperature
difference at the ends of the wire and the voltage caused by it.
Again, the function $F_1(x)$ falls off exponentially at $|x|\gg 1$,
see Fig.~\ref{fig:fig2}, and its analytical form is given by
Eq.~(\ref{eq:Fn}). Since $S$ is the leading contribution, it is a
direct measure of the inter-channel interaction. Furthermore, since
$S$ and $\delta G$ are functions of $(\eF-\ern)/\kb T$, measurements
performed at different temperatures are predicted to collapse to the
two curves of Fig.~\ref{fig:fig2}, when properly scaled. The shape
of the curve $F_1(x)$ and in particular the sign change can be
understood in terms of allowed inter-channel relaxation processes,
as explained and illustrated in Fig.~\ref{fig:fig3}.

A remarkably good estimate, $F_1(x)\approx (\pi^2/3)dF_0/dx$, can
be obtained from the Mott's formula~\cite{Mott1936},
\begin{align}\label{eq:mott-law}
S^{\textrm{M}}=\frac{\pi^2}{3} \frac{\kb}{e} \kb T \frac{1}{G}
\frac{\dd G}{\dd \eF},
\end{align}
which is expected to be a good approximation at low temperatures for
non-interaction electrons \cite{lunde05}. There is no \emph{a priori}
reason for Mott's formula to work here, as we consider effects of
inelastic scattering. We therefore stress that interaction effects can
be seen in an experiment even without a manifesting violation of the
Mott formula. (We note though, that violation of the Mott formula has
been detected in experiments~\cite{appleyard00}.)

\begin{figure}
\centerline{\includegraphics[width=0.4\textwidth]{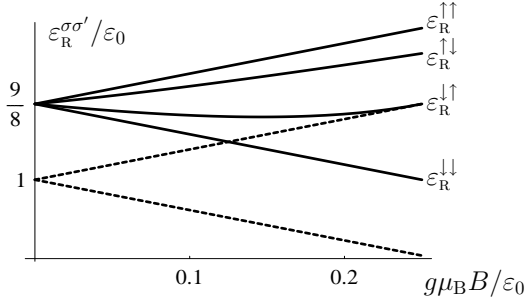}}
\caption{(Color online) Splitting of the resonance features in
conductance and thermopower into four features as a function of
magnetic field $B$. Dashed lines represent the evolution of the
positions of single-particle peaks in the thermopower.
\label{fig:fig4}}
\end{figure}

Magnetic field introduces Zeeman splitting of the electron states.
This, in turn, splits the single resonance for both conductance and
thermopower into four resonances, see Fig.~\ref{fig:fig4}. Two of
the resonances correspond to the transitions involving electrons
with the \emph{same} spins. With the increase of the magnetic field
$B$, splitting in this doublet, in terms of $\eF$, equals $g\mub B$
and coincides with the Zeeman splitting of the quantized steps in
the conductance (here $g$ and $\mub$ are the electron $g$-factor in
a quantum wire and Bohr magneton, respectively). The amplitude of
the resonances involving electrons with the same spin is smaller
than the one given by Eqs.~(\ref{eq:conductance}) and
(\ref{eq:thermopower}) by a parameter $\propto (\kb T/\eF)^2$. This
suppression is a manifestation of the Pauli principle: at $T=0$ the
scattering process we consider would involve two electrons in the
same orbital and spin state, see Fig~\ref{fig:fig1}. The full form
of the conductance correction and the thermopower in this doublet is
given below in Eq.~(\ref{eq:dG-S-same-spin}). The two remaining
resonances form another doublet with splitting $g\mub B/2$ at small
fields, see Eq.~(\ref{eq:erssbar}) and Fig.~\ref{fig:fig4} for
details. These resonances correspond to the transitions involving
electrons with \textit{opposite} spins and are therefore not
suppressed by the Pauli principle; Eqs.~(\ref{eq:conductance}) and
(\ref{eq:thermopower}) can be used for an estimate of the amplitude
of these two resonances. However, as seen on Fig.~\ref{fig:fig4},
the position of the lower of these resonances approach the bottom of
the spin-split band, $\e_0+g\mub B/2$, which tends to mask the
interaction-induced structure.

We thus find that the interaction-induced features in transport
properties scale with the temperature as $T$ or $T^3$. They are
associated with electron scattering at energies close to the Fermi
level. This should be compared to the situation without interactions
where finite-temperature corrections to the quantized conductance
and to the zero value of $S$ are of the order of $\exp(-\e_0/8\kb
T)$, whichs stems from the exponentially small probability of having
holes at the bottom of band $2$. Thus at sufficiently low
temperatures the features we consider are dominant.

Next, we outline the derivation of the electron-electron interacition
effects on the current. We start from the Boltzmann equation,
\begin{align}
v_{kn}\partial_{x}f_{kn}(x)=\mathcal{I}_{knx}[f],
\label{eq:Boltzmann-eq}
\end{align}
where $f_{kn}(x)$ is the distribution function, $n$ the channel
index, $v_{kn}=\frac{1}{\hbar}\partial_k\e_{kn}$ the velocity, and
$\mathcal{I}_{knx}[f]$ is the electron-electron collision
integral. The reflectionless contacts of the quantum wire
\cite{glazman88} and the shift in chemical potential and
temperatures \cite{kdot} are introduced by the boundary conditions
\begin{subequations}
\begin{align}
f_{kn}(x=0)&= f^0_{\textrm{L},kn}\quad \textrm{for}\quad k>0,\\
f_{kn}(x=L)&= f^0_{\textrm{R},kn}\quad \textrm{for}\quad k<0,
\end{align}
\end{subequations}
where $f_{\textrm{L}/\textrm{R},kn}^0$ is the Fermi distribution
for the left and right lead, respectively, including their
temperatures $T_{\textrm{L}/\textrm{R}}$ and chemical potentials
$\mu_{\textrm{L}/\textrm{R}}$. In the absence of interactions
($\mathcal{I}_{knx}[f]=0$), the solution to the Boltzmann equation
is simply given by
\begin{align}
f_{kn}^{(0)}= f^0_{{L,kn}}\theta(k)+f^0_{{R,kn}}\theta(-k).
\label{eq:f-0}
\end{align}
Interaction between electrons yield the collision integral
\begin{align}
\mathcal{I}_{k_1^{}n_1^{}x}[f]=&-\!\!\sum_{\substack{\sigma_2^{}\\\sigma_{1^{\prime}}\sigma_{2^{\prime}}}}
\sum_{\substack{n_2^{}\\n_{1^{\prime}}n_{2^{\prime}}}}
\sum_{\substack{k_2^{}\\k_{1^{\prime}}k_{2^{\prime}}}}\!
W_{12;1^{\prime}2^{\prime}} \big[f_1f_2(1-f_{1^{\prime}})\nonumber\\
&\times(1-f_{2^{\prime}})-f_{1^{\prime}}f_{2^{\prime}}(1-f_1)(1-f_2)\big],
\label{eq:collision-integral-2-particle-full}
\end{align}
where $f_i=f_{k_in_i}(x)$, and primes denote electron states after the
collision. The scattering rate $W_{12;1^{\prime}2^{\prime}}$ is found
using the Fermi Golden rule.  To find the distribution, we expand it
powers of $W_{12;1^{\prime}2^{\prime}}$ as $f=f^{(0)}+f^{(1)}+\cdots$.
To the leading order, we can therefore insert $f^{(0)}_{kn}$ from
Eq.~(\ref{eq:f-0}) into the collision integral in the right-hand-side
of Eq.~(\ref{eq:Boltzmann-eq}) to get
\begin{equation}
f_{kn}^{(1)}(x)=\left[\frac{x}{v_{kn}}\,\theta(k)+
\frac{x-L}{v_{kn}}\,\theta(-k)\right]\mathcal{I}_{kn}[f^{(0)}].
\label{eq:Solution-to-BE-f-1}
\end{equation}
The electric current now follows as (with $e>0$)
% \begin{align}
% I&=\frac{(-e)}{L}\sum_{\sigma n k}v_{kn} f^{(0)}_{kn}
% -(-e)\sum_{\sigma n k<0}\mathcal{I}_{kn}[f^{(0)}]\nonumber\\
% &=I^{(0)}+I^{(1)}, %\label{eq:current-first-order-in-W}
% \end{align}
\begin{eqnarray}
&I&=I^{(0)}+I^{(1)}\,,
\label{eq:current-first-order-in-W}\\
&I^{(0)}&=\frac{(-e)}{L}\sum_{\sigma n k}v_{kn}
f^{(0)}_{kn}\,\,,\,\, I^{(1)}= e\!\!\sum_{\sigma n
k<0}\mathcal{I}_{kn}[f^{(0)}],\nonumber
\end{eqnarray}
where $I^{(0)}$ corresponds to the Landauer formula for fully open
channels, and the term $I^\mathrm{(1)}$ is the \emph{correction to
the current due to interactions}. To calculate $I^\mathrm{(1)}$,
we linearize $\mathcal{I}_{kn}[f^{(0)}]$ in
$eV=\mu_{\textrm{L}}-\mu_{\textrm{R}}$ and $\Delta
T=T_{\textrm{R}}-T_{\textrm{L}}$ and divide the summation over
quasi wave vectors into positive and negative values using
$W_{12;1^{\prime}2^{\prime}}=W_{21;1^{\prime}2^{\prime}}=W_{1^{\prime}2^{\prime};12}$,
etc. \cite{LFG}. We obtain
\begin{align}
I^\mathrm{(1)}=2&(-e)\!\!\!\sum_{\substack{\sigma_1^{}\sigma_2^{}\\\sigma_{1^{\prime}}\sigma_{2^{\prime}}}}
\sum_{\substack{n_1^{}n_2^{}\\n_{1^{\prime}}n_{2^{\prime}}}}
\sum_{\substack{k_1^{}<0,k_2^{}>0\\k_{1^{\prime}}>0,k_{2^{\prime}}>0}}
\!\!\!\!\!\!
W_{12;1^{\prime}2^{\prime}}f_1^0f_2^0(1-f_{1^{\prime}}^0)\nonumber\\
&\times (1-f_{2^{\prime}}^0)\left[\frac{\Delta T}{\kb
T^2}(\e_1-\mu)-\frac{eV}{\kb T}\right], \label{eq:current-int}
\end{align}
where $f^0_i=1/[\exp((\e_{k_in_i}-\mu)/\kb T))+1]$ is the Fermi
function. For two channels, we find that the only combination of
channel indices, which gives a contribution non-exponential in
temperature, is $n_2^{}=1$ and $n_1^{}=n_1^{\prime}=n_2^{\prime}=2$.
Moreover, this combination is non-exponential only if the Fermi
energy is within a range of $\kb T$ from $\ern$. An important point
of the result (\ref{eq:current-int}) is that the number of left and
right moving electrons has to change in order for the current to
change. In essence, this is due to the cancellation of the velocity
in the distribution function, Eq.~(\ref{eq:Solution-to-BE-f-1}), and
in the current definition, Eq.~(\ref{eq:current-first-order-in-W}).
In fact, the cancellation can be shown to be valid to all orders in
the interaction \cite{LFG}.  A similar situation occurs for the
Coulomb drag response \cite{gurevich98,mortensen01} of mesoscopic
structures.

At low temperatures, $T\ll (\eF-\e_0)/\kb$, we can now find the
thermopower and conductance for both zero and nonzero magnetic
field. From Eq.~\eqref{eq:current-int}, we obtain the linear
response current as $I=(G_T^{(0)}+G_T^{(\textrm{1})})\Delta
T-(G^{(0)}+G^{(\textrm{1})})V$, where $G$ is the conductance and
$G_T$ is the thermo-electric coefficient. Furtermore, since
$G_T^{(0)}\propto e^{-(\eF-\e_0)/\kb T}$ at the plateaus
\cite{vanhouten92} and $G_T^{(\textrm{1})}\propto T$ or $T^3$ one
obtains at low temperatures the thermopower $S\simeq
G_T^{(\textrm{1})}/G^{(0)}$. Moreover, the contribution to lowest
order in temperature is found by linearizing the quadratic
dispersion and using a constant interaction in Fourier space in
Eq.~(\ref{eq:current-int}). For $\eF$ near $\ern$, we finally
obtain after some algebra the zero-magnetic-field conductance
correction, $\delta G=G^{(\textrm{int})}$, and thermopower given
in Eqs.~\eqref{eq:conductance} and \eqref{eq:thermopower} (plotted
in Fig.~\ref{fig:fig2}). The dimensionless scaling functions $F_0$
and $F_1$ are determined by
\begin{subequations}
\label{eq:Fn}
\begin{align}
F_n^{}(x)&=\int_{-\infty}^{\infty} \dd z z^n h(z,x),\\
h(z,x)&=\frac{-(2x+z)}{4\sinh(2x+z)\cosh(\frac{z}{2})\cosh(2x+\frac{3z}{2})}.
\end{align}
\end{subequations}
Further, in Eqs.~\eqref{eq:conductance} and \eqref{eq:thermopower}
we have defined the effective inter-channel electron-electron
scattering length as
\begin{equation}
    \lee=
\frac{2}{27}
\frac{2\pi}{k_{\textrm{F}1}}
\left(\frac{\hbar
v_{\textrm{F}1}}{|V_{2k_{\textrm{F}2}}^{2122}|}\right)^2,
\label{eq:leedef}
\end{equation}
where $V_{2k_{\textrm{F}2}}^{2122}$ is the electron-electron
interaction at twice the Fermi vector for the second channel ({\it
  i.e.,} upper sub-band) $2k_{\textrm{F}2}$. Both the curvature of the
electron dispersion relation and the momentum dependence of the
interaction give rise to small corrections, {\it e.g.},
$S(\eF=\ern)\propto T^2$.  These corrections to higher order in
temperature shift the point where the thermopower coefficient
changes sign, but the overall behavior is still the same.

Magnetic field splits the resonance at $\eF=\ern$ into four points
corresponding to scattering between electrons with identical or
different spins. For scattering between electrons with identical
spins, momentum and energy conservation lead to resonances occurring
at ($\sigma=\pm 1= \uparrow\downarrow$)
\begin{equation}\label{eq:erss}
\er^{\sigma\sigma}=\frac98 \e_0+\sigma\frac12 g\mub B.
\end{equation}
However, the features which originate  from scattering between
different spins split up in a different way
($\bar{\sigma}=-\sigma$),
\begin{equation}\label{eq:erssbar}
\er^{\sigma\bar{\sigma}}= \e_0 \frac{9+20\sigma g\mub
B/\e_0+8(g\mub B/\e_0)^2}{8(1+2\sigma g\mub B/\e_0)}.% \theta(\e_0/4+\sigmag\mub B)
\end{equation}
For $g\mub B>\e_0/4$ only three resonances remain, since the
scattering at $\eF=\er^{\downarrow\uparrow}$ can no longer conserve
both momentum and energy (similarly when $g\mub B<-\e_0/4$ the
$\eF=\er^{\uparrow\downarrow}$ resonance is absent).

In the regime $\kb T\ll g\mub B\ll\eF-\e_0$, the thermopower and
conductance change around each of the points
$\eF=\er^{\sigma^{}\sigma^{\prime}}$ can be found separately. The
 matrix element of the interaction potential
\begin{align}
\langle &
k_{1^{\prime}}2\sigma_{1^{\prime}}k_{2^{\prime}}2\sigma_{2^{\prime}}|V|k_{1^{}}2
\sigma_1^{}k_{2^{}}1\sigma_2^{}\rangle=
\frac{\delta_{k_{1}^{}+k_{2}^{},k_{1^{\prime}}^{}+k_{2^{\prime}}^{}}
}{L}\nonumber\\
&\times\left[V_{k_{1^{\prime}}^{}-k_1^{}}^{2122}
\delta_{\sigma_{1^{}}^{},\sigma_{1^{\prime}}}\delta_{\sigma_{2^{}}^{},\sigma_{2^{\prime}}}
- V_{k_{2^{\prime}}-k_{1^{}}^{}}^{2122}
\delta_{\sigma_{1^{}}^{},\sigma_{2^{\prime}}}\delta_{\sigma_{2^{}}^{},\sigma_{1^{\prime}}}
\right], \label{eq:Coulomb-matrix-element}
\end{align}
consist of a direct (first) and exchange (second) term leading to
the remarkable difference between the resonance amplitude at
$\er^{\sigma\bar{\sigma}}$ and $\er^{\sigma\sigma}$. The direct
and exchange contributions to the scattering amplitude cancel each
other for a point-like interaction and equal spins (Pauli
principle). Finite contributions to the transport characteristics
appear due to the momentum dependence of the interaction matrix
elements (\ref{eq:Coulomb-matrix-element}) and are of higher power
in temperature:
\begin{subequations}
\begin{align}
S_{\sigma\sigma}&=\frac{\kb}{e} \frac{6L}{\lee'} \bigg(\frac{\kb
T}{\eF}\bigg)^3\,\widetilde{F}_1^{}\!\left(\frac{\eF-\er^{\sigma\sigma}}{\kb
T}\right),\\
\delta G_{\sigma\sigma}&=\frac{4e^2}{h}
\frac{6L}{\lee'}\bigg(\frac{\kb T}{\eF}\bigg)^3\,
\widetilde{F}_0^{}\!\left(\frac{\eF-\er^{\sigma\sigma}}{\kb
T}\right),
\end{align}\label{eq:dG-S-same-spin}
\end{subequations}
where
\begin{equation}\label{eq:leem}
    \lee'=\frac{2}{27}\frac{2\pi}{k_{\mathrm{F}1}}\left(\frac{\hbar v_{\mathrm{F}1}^{\sigma}}
    {k_{\mathrm{F}1}^{\sigma}|\partial_q
    V_{2k_{\mathrm{F}2}^\sigma}^{2122}|}\right)^2.
\end{equation}
The new dimensionless functions $\widetilde{F}_0^{}$ and
$\widetilde{F}_1^{}$ are defined as
\begin{equation}
\widetilde{F}_n^{}(x)=\frac{1}{16}\int_{-\infty}^{\infty}\!\!\!\!\!\dd
z z^n h(z,x)\left[\frac{4\pi^2}{3}+\frac{1}{3}(4x+2z)^2\right],
\end{equation}
being well approximated by $F_0$ and $F_1$ in Fig.~\ref{fig:fig2}.
The predicted evolution of the resonance structure with the increase
of magnetic field should be a good candidate for experimental
verification of the interaction effects.

In summary, we found interaction-induced resonance points in the
Fermi energy in the conductance and thermopower $S$ of a short clean
multi-channel quantum wire adiabatically connected to leads. The
resonances in thermopower offer a way to observe the effect of
inter-channel electron-electron interaction and to measure its
strength. We found $S$ to be dominated entirely by interactions for
$|\eF-\ern|/\kb T\lesssim 5$ at low temperatures and to have the
scaling form presented in Eq.~(\ref{eq:thermopower}). Furthermore,
the thermopower coefficient and the conductance develop distinct
features in a finite magnetic field.

\emph{Acknowledgements} -- The work at the University of Minnesota
is supported by NSF grants DMR 02-37296 and DMR 04-39026. We thank
H. Buhmann, M. Garst, A.-P. Jauho, M.  Khodas, L. W.  Molenkamp,
M. Pustilnik and H. Smith for discussions. The hospitality of
MPIPKS-Dresden (L.I.G.) and of the William I. Fine Theoretical
Physics Institute, University of Minnesota (A.M.L.) is gratefully
acknowledged.


\begin{thebibliography}{10}
\bibitem{kdot}
The term $\dot{k}\partial_kf$ is not present in the Boltzmann
equation, since
  the bias and temperature difference are included in the boundary condition.
\end{thebibliography}
\end{document}